\documentclass[a4paper]{EslabStyle}
\usepackage{graphics}

\begin{document}

\title{Cosmic Star Formation Rate from Gamma-Ray Attenuation}

\author{T. M. Kneiske \and K. Mannheim} \institute{Universit\"ats Sternwarte G\"ottingen, Germany}

\maketitle 

\begin{abstract}
We compute the attenuation of gamma-rays from remote quasars due to interactions with ambient 
low-energy photons produced by stars in galaxies and show that the next-generation gamma ray 
telescopes can probe the cosmic star formation rate at high redshifts.

\end{abstract}

\section{Introduction}

The cosmic star formation rate (CSFR) at high redshifts is a key probe 
to determine the time-table of structure formation in the universe
and hence to further constrain cosmology.  Number counts of galaxies
only give access to small parts of the sky and are subject to severe
selection effects depending on the wavelength range of the observational
campaign such as, for example, extinction by dust using deep photometry in 
the optical.  Furthermore, absolute sky photometry is biased due to
galactic foregrounds making it difficult to assess truly diffuse
extragalactic background light (EBL) which could arise due to the decay
of unstable relics from the early universe.  Therefore, an independent method 
unaffected by these obstacles is required, and we suggest along with 
Salamon \& Stecker (1998) and Primack et al. (1999) to use the
attenuation of gamma rays from distant quasars for this purpose.

EGRET has detected that many flat-spectrum radio quasars at high redshifts 
emit gamma rays at least up to 10~GeV showing generally no indication
for a spectral break.  A few nearby flat-spectrum radio sources have
even been detected up to TeV ($10^{12}$~eV) energies using ground-based Cherenkov
telescopes suggesting that at least some of the sources exhibit
smooth power-law spectra in the GeV-to-TeV energy range.
The gamma rays of energy $E_\gamma$ can interact with ambient low-energy photons 
of energy $\epsilon$ creating 
electron-positron pairs provided the threshold energy $E_\gamma\sim (m_ec^2)^2/\epsilon$ 
is exceeded.  As a result, the gamma ray sources should show a cutoff in their
spectrum at the energy where the mean free path for pair creation
becomes shorter than the distance to the sources, i.e. where $\tau_{\gamma\gamma}(E_\gamma)
\sim \sigma_{\gamma\gamma}n_{\rm ebl}[(m_ec^2)^2/E_\gamma]d_z\ge 1$ where
$\sigma_{\gamma\gamma}\simeq \sigma_{\rm T}/3$, $n_{\rm ebl}$ denotes the
photon number density of the EBL, and $d_z$ is the cosmological distance of
the gamma ray source.  Generally, since the number density of background
photons increases with decreasing photon energy, and thus increasing gamma
ray energy above threshold, the more distant gamma ray sources must show lower
cutoff energies than the closer gamma ray sources and the precise running of
the cutoff energies with redshift will tell us about the shape and evolution
of the EBL.  

Our strategy now is to develop a simple model for the redshift-dependent EBL based on the observed
(and extinction-corrected) CSFR and then to determine the corresponding values of gamma ray
cutoff energies.  We start with a description of the galaxy evolution templates used
to determine the EBL when folded with the CSFR, and proceed with the computation of
the gamma ray optical depth. Finally, we discuss that
possible deviations of future measurements
of the gamma ray cutoff energies would require changes in
the EBL model (e.g. in the CSFR) or force us to consider additional sources to contribute to the EBL.

\section{Spectral energy distribution}
\label{sec:SED}

\begin{figure}[ht]
\resizebox{\hsize}{!}{\includegraphics{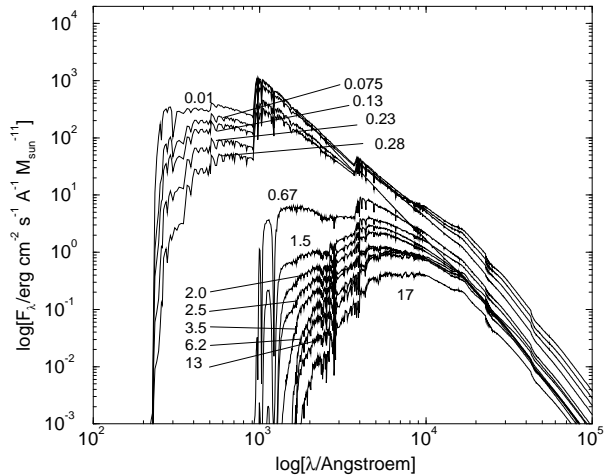}}
\caption{Spectral energy distribution of a simple stellar population with mass $10^{11}M_\odot$ 
and solar metalicity evolving from an age of 0.01 Gyrs to 17 Gyrs. \label{fig1}}
\end{figure}
\begin{figure}[ht]
\resizebox{\hsize}{!}{\includegraphics{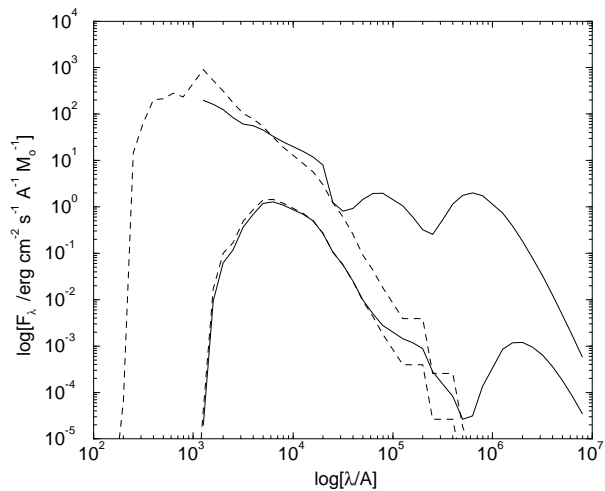}}
\caption{Spectral energy distributions of a simple stellar population for the
early (the first 0.28 Gyrs) and late (after 0.28 Gyrs) phases.
The solid line includes (dashed line excludes) absorbtion and reemission of the ISM.}\label{figneu}
\end{figure}

We use the spectral synthesis model calculated by M\"oller et al. (1999), which describes the evolution of 
the spectral energy distribution (SED) of a single burst population (SSP) with time
(see Figure~\ref{fig1}). The stars have solar metalicity and Salpeter IMF.
We simplify the temporal evolution by reducing it to only two spectral states.
The first state lasts from ignition to an age of 0.28 Gyrs. We call this state the 
\emph{early phase} and it is dominated by the ultraviolet and optical radiation of 
young massive stars. 
The second state is defined to begin at an age of 0.28 Gyrs. 
We call this state the \emph{late phase} and it is dominated by the optical and infrared light of the 
older stars in the population.
To get the spectrum representative for each phase we calculate the corresponding mean SED from the
templates. Note that the late phase lasts over a cosmologically significant portion of
time and therefore must be weighted with the CSFR.
Furthermore,
we include the effect of the interstellar medium (ISM) by 
assuming an isotropic and homogeneous gas and dust distribution around the stars of the SSP.
We use Osterbrook's Case B for optical thick gas clouds, i.e. total absorbtion of photons below the
wavelength of 911 {\AA} and reemission of the absorbed power in the optical regime
by bremsstrahlung and $H_\alpha$ line emission.
The dust is treated the same way as in Desert et al. (1990) considering
a three component model with big grains, small grains and PAH's. 
For the absorbtion we use an extinction curve similar to the one of the SMC. 
The reemission is described by four Planck spectra with two temperatures for each phase 
reprensenting warm and cold dust components. 
The resulting spectra including absorbtion and re-emission are shown in figure~\ref{figneu}. 

\section{Cosmic star formation rate}
\label{sec:csr}

\begin{figure}[ht]
\resizebox{\hsize}{!}{\includegraphics{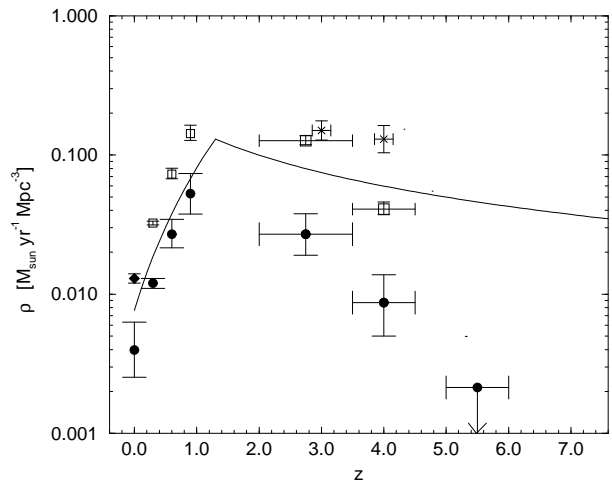}}
\caption{Cosmic star formation rate $\stackrel{\cdot}{\rho}_{\ast}$ vs. redshift $z$.
The data are taken from Madau (1996; solid circles), 
corrected by Pettini et al. (1998; open squares), 
Gallego (1995; diamond) and further data from Steidel et al. (1999; stars). 
The solid line indicates
the fit used in our model \label{fig2}}
\end{figure}

Now let us look at galaxy evolution on global scales.
In our model, the stars in galaxies are all represented by SSPs.
So far we did not include a particular 
star formation rate, and we will do this now by using the cosmic star formation rate (CSFR)
as determined from galaxy counts in the optical (e.g., Madau 1996).
Since dust and gas absorbs optical photons, the original Madau CSFR underestimates the
true SFR in particular at high redshifts by at least a factor of
$\sim 3$ (Pettini et al. 1998) or maybe even $\sim 4.7$ (Steidel et al. 1999). 
All data agree, however, in a CSFR increasing up to a redshift 
$z=z_{\rm b}\sim1-2$ and decreasing at $z>z_{\rm b}$, see Fig.~\ref{fig2}.
We adopt in the following a power-law fit to this CSFR shown as the solid line
in Fig.~\ref{fig2}
\begin{equation} \dot{\rho_\ast}(z) \propto (1+z)^\alpha \end{equation}
with $ \alpha=\alpha_m>0$ for $z < z_b$ and $\alpha=\beta_m<0$ for $z > z_b$. 

\section{Emissivity}
\label{sec:emi}
\begin{figure}[ht]
\resizebox{\hsize}{!}{\includegraphics{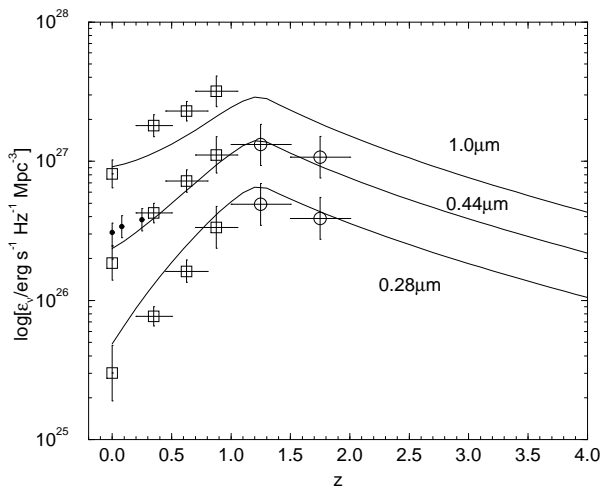}}
\caption{Emissivity vs. redshift . Data taken from Lilly et al. (1996; squares), Conolly et al. (1997; circles) and Ellis et al. (1996; dots).\label{fig3}}
\end{figure}

Solving the convolution integral 

\[\mathcal{E}_{\lambda}(t) = \int_0^t \mathcal{F}_{\lambda}(t-t')\stackrel{\cdot}{\rho_\ast}(t') dt'\]
we can calculate the emissivity at different cosmic times $t=t(z)$.
For the comparison with measurments we will integrate over redshift $z$ and change to the 
\emph{comoving} emissivity

\begin{equation}\mathcal{E}_{\lambda}(z) = (1+z)^{-2} \int_z^{z_{m}} \mathcal{F}_{\lambda}(t(z)-t(z'))\stackrel{\cdot}{\rho}_{\ast}(z') \frac{dt'}{dz'} dz'. 
\label{eq:emislambda}
\end{equation}
We used the parameters in Tab.~1 to obtain the results shown in Fig.~\ref{fig3}. 
The emissivity $\mathcal{E}_{\lambda}(z)$ is plotted for three different wavelengths
demonstrating satisfactory agreement. 
Note that in particular the steeper increase at 0.28 $\mu$m and the more
shallow increase at 1.0 $\mu$m are well reproduced by the model.

\begin{table}[h]

\begin{center}
\begin{tabular}{c c c c c c}
\hline 
$E_{B-V}^e$	&$E_{B-V}^l$	&$\alpha_m$	&$\beta_m$	&$z_b$	&$y_{m} [M_{\odot} yr^{-1} Mpc^{-3}]$	\\ \hline
0.1	&0.03	&3.4	&-2	&1.3	&0.13	\\
\hline 

\end{tabular}
\caption{Choosen parameter for the calculation of the emissivity. Note that we used the better value $\beta_m=-1$ for the CSFR (Fig. 3) and the background (Fig. 5) calculation.}
\end{center}
\label{tab:parameter}
\end{table}

\section{Extragalactic background light}

The intensity of the EBL follows from the emissivity

\begin{equation}
\lambda I_{\lambda} = \lambda \frac{c}{4\pi} \int_z^{z_m}  \mathcal{E}_{\lambda'}(z) \frac{dt'}{dz'} \left( \frac{1+z}{1+z'} \right)^2  dz'.
\label{eq:hinter}
\end{equation}

\begin{figure}[ht]
\resizebox{\hsize}{!}{\includegraphics{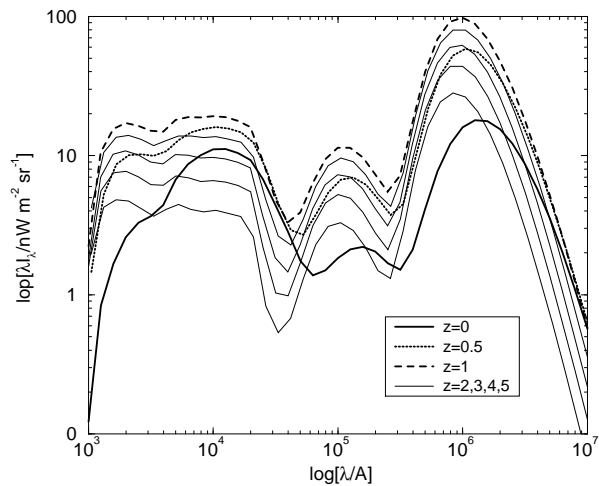}}
\caption{Diffuse Extragalactic background light due to stars in galaxies, for various redshifts.\label{fig4}}
\end{figure}

\begin{figure}[ht]
\resizebox{\hsize}{!}{\includegraphics{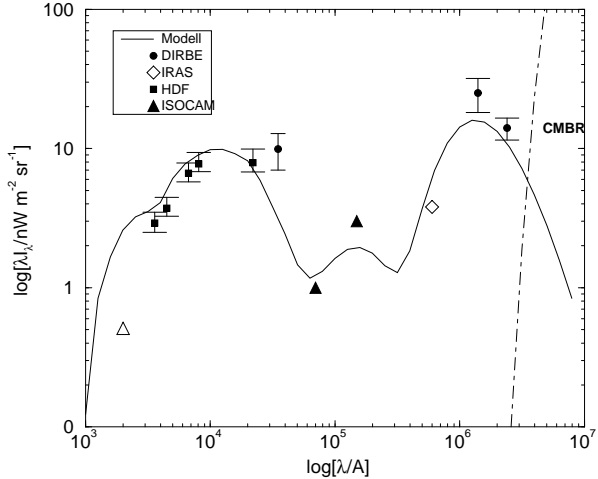}}
\caption{Comparison of the model with the observed present-day EBL. Data points refer to
Armand et al. (1994; open triangle), Pozetti et al. (1998; squares/HST), Dwek et al. 
(1998; filled circle at 35$\mu$m/DIRBE), Altieri et al. (1998; triangle/ISOCAM), and Hauser et al. (1998;
diamond/IRAS, blue filled circles/DIRBE).
\label{fig5}}
\end{figure}

We compute the background intensity at several redshifts and show the results in Fig.~\ref{fig4}. 
Beyond a redshift of $z\sim 1.3$, 
the EBL decreases because of the decreasing CSFR and the expansion of
the universe.
The shape of the spectrum at high redshifts is determined by the \emph{early phase} spectrum,
whereas 
for lower redshifts the \emph{late phase} becomes more and more important.  This is a consequence of 
of the maximum star formation rate at redshifts $z=1-2$ from which the stellar populations
have considerably aged until the present.
There are no direct measurements of the EBL at high redshift (with the exception of inferences of
the UV background based
on the proximity effect which are consistent with our model predictions if we employ empirical
power law extensions of the UV spectra beyond the Lyman edge). 
We can, however, compare our results with the present-day data obtaining
good agreement as shown in Fig.~\ref{fig5}.

\section{Gamma ray attenuation}
The gamma-ray attenuation can be parametrized by the optical depth for photon-photon pair production 

\begin{eqnarray}
\tau(E_\gamma,z_q)=c\int_{0}^{z_q} \int_{0}^{2} \int_{\epsilon_{gr}}^{\infty} \frac{dl}{dz'}  \frac{\mu}{2}  
n_{\rm ebl}(\epsilon,z') (1+z)^3\cdot \\ \cdot\sigma_{\gamma\gamma}(E_\gamma,\epsilon,\mu, z')  d\epsilon\  d\mu\ dz'\nonumber
\end{eqnarray}

\begin{figure}[ht]
\resizebox{\hsize}{!}{\includegraphics{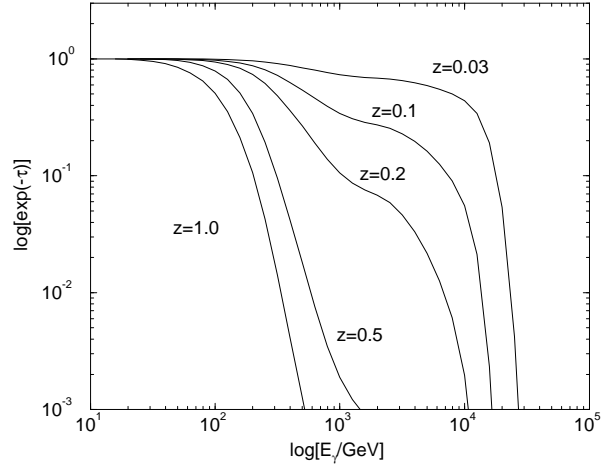}}
\caption{Attenuation factor $\exp[-\tau(E_\gamma,z_q)]$ for gamma-rays as a function of gamma ray energy 
$E_\gamma$ for various quasar redshifts $z_q$.
\label{fig6}}
\end{figure}

Using Eq.~(3) for the intensity of the background photons we can calculate the number density 
$n_{\rm ebl}(\epsilon, z')$. Inserting $n_{\rm ebl}(\epsilon, z')$ in Eq.~(4) we can compute the 
optical depth for various redshifts. The optical depth enters the attenuation factor 
$\exp[-\tau(E_\gamma,z_q)]$ for gamma rays (see Fig.~\ref{fig6}) modifying quasar power law
spectra.  Future gamma ray observations
with ground-based air Cherenkov telescopes in the energy range 10~GeV to 10~TeV
can efficiently probe  the run of the CSFR with redshift.  E.g., the faster the decrease
above the maximum of the CSFR (the larger the slope $\beta_m$ in Eq.~(1)), the less
attenuation occurs for quasars at redshifts $z\gg 1$.  On the other hand, a plateauish
CSFR combined with the effect of dust keeps the UV radiation density low at high redshifts,
but leads to a strong FIR bump in the EBL which leads to strong attenuation of 
$>$ 10 TeV gamma rays
from nearby $z\ll 1$ radio sources.


\section{Summary and conclusions}
In order to facilitate a comparison of gamma ray cutoff energies in the
spectra of high-redshift quasars with theoretical predictions of gamma ray attenuation
due to pair creation, we developed a simple model for the evolving EBL based on the observed
CSFR taking into account extinction and re-emission by dust and gas. Further model inputs
are the IMF, the average metalicity, relative contributions of warm and cold dust to
the emission of early- and late-type galaxies, and the Hubble constant for which we have
adopted values in close agreement with observations.
The CSFR at high redshifts is not accurately determined by the data, and we have tried
to use a value which is consistent with both the present-day EBL and high-redshift galaxy counts.
So far we have neither considered the extension of the galaxy spectra beyond the
Lyman edge, the metalicity dependence of galaxy spectra, 
nor the contribution of quasars to the EBL.  The predicted attenuation of gamma rays
from nearby radio sources such as Mrk501 or Mrk421 leads to a quasi-exponential cutoff near
10 TeV due to interactions with the infrared part of the EBL,  and to a cutoff near 100 GeV
due to interactions with the optical/UV part of the EBL for sources with redshifts in excess
of one.

\end{document}